\documentstyle[prl,aps,amsfonts,preprint,epsfig,floats]{revtex}

\tighten

\newcommand{\be}{\begin{eqnarray}}
\newcommand{\ee}{\end{eqnarray}}

\newcommand{\fract}[2]{{\textstyle\frac{#1}{#2}}}

\newcommand{\lapeq}{\stackrel{\scriptscriptstyle\raisebox{-3.0mm}{$<$}}
{\scriptscriptstyle \raisebox{-1.5mm}{$\sim$}}}
\newcommand{\gapeq}{\stackrel{\scriptscriptstyle\raisebox{-3.0mm}{$>$}}
{\scriptscriptstyle \raisebox{-1.5mm}{$\sim$}}}

\newcommand{\Tr}{\mbox{Tr}}
\newcommand{\skp}{\epsilon}

\begin{document}

\title{Hyperon Beta--Decay and Axial Charges of the Lambda\\
in view of Strongly Distorted Baryon Wave--Functions}
\author{H.~Weigel{}\footnote{Heisenberg Fellow}
\footnote{e-mail:herbert.weigel@uni-tuebingen.de}}

\address{{~}\\Center for Theoretical Physics, Laboratory for
	Nuclear Science
	and Department of Physics \\
	Massachusetts Institute of Technology,
	Cambridge, Massachusetts 02139\\
        and\\
        Institute for Theoretical Physics, T\"ubingen University\\
        Auf der Morgenstelle 14, D--72074 T\"ubingen, Germany}

\maketitle

\begin{abstract}
Within the collective coordinate approach to chiral soliton 
models we suggest that breaking of $SU(3)$ flavor symmetry mainly 
resides in the baryon wave--functions while the charge operators 
maintain a symmetric structure. Sizable symmetry breaking in the 
wave--functions is required to reproduce the observed spacing in the 
spectrum of the $\fract{1}{2}^+$~baryons. The matrix elements of the
flavor symmetric charge operators nevertheless yield $g_A/g_V$ ratios 
for hyperon beta--decay which agree with the empirical data approximately 
as well as the successful $F$\&$D$ parameterization of the Cabibbo 
scheme. Demanding the strangeness component in the nucleon to vanish 
in the two flavor limit of the model, determines the structure of the 
singlet axial charge operator and yields the various quark flavor components 
of the axial charge of the $\Lambda$--hyperon. The suggested picture 
gains support from calculations in a realistic model using 
pion and vector meson degrees of freedom to build up the soliton.
\end{abstract}

\bigskip

\section{Introduction}

The study of the axial current matrix elements (or quark spin 
structure) of the~$\Lambda$ hyperon, which is an interesting subject 
in its own, has gained further attraction as it has been 
suggested \cite{Ja96} that polarized $\Lambda$'s 
could be utilized to gain information about the proton spin 
structure, {\it i.e.} the nucleon axial vector matrix elements. For 
this to be a sensible program it is necessary that large polarizations 
of the {\it up} and {\it down} quarks, $\Delta U_\Lambda$ and 
$\Delta D_\Lambda$, in the iso--singlet $\Lambda$ carry over to the 
corresponding fragmentation functions. Although this is an 
assumption one would expect the {\it non--strange} $\Lambda$ 
fragmentation functions to be small if model calculations undoubtly showed 
that $\Delta U_\Lambda$ and $\Delta D_\Lambda$ were small. On the
other hand, if model calculations indicated that $\Delta U_\Lambda$ and 
$\Delta D_\Lambda$ were large, it would provide sufficient motivation 
to study and measure these fragmentation functions~\cite{Bu93,Fl97}.
Using results on the axial current matrix elements from deep--inelastic
scattering as well as hyperon beta--decay data together with flavor 
covariance indeed results in sizable polarizations for the {\it non--strange} 
quarks, $\Delta U_\Lambda=\Delta D_\Lambda\approx-0.20$ together with 
$\Delta S_\Lambda\approx0.60$ for the {\it strange} 
quark~\cite{Ja96,Bu93,Bo98}. The use of flavor covariance is 
motivated by the feature that the 
Cabibbo scheme \cite{Ca63} utilizing the $F$\&$D$ parameterization for 
the flavor changing axial charges works unexpectedly well~\cite{Fl98} as 
the comparison in table~\ref{empirical} exemplifies. In the present note
we will study an approach which allows the incorporation of deviations
from the flavor symmetric formulation, after all $SU(3)$--flavor is
not an exact symmetry. Clearly, any model that reproduces the data
equally well as the Cabibbo scheme with a minimal set of parameters can
be regarded as a reasonable description of hyperon beta--decay.

\begin{table}[h]
\caption{\label{empirical}\sf The empirical values for the 
$g_A/g_V$ ratios of hyperon beta--decays \protect\cite{DATA}, 
see also~\protect\cite{Fl98}. 
For the process $\Sigma\to\Lambda$ only $g_A$ is given.
Also the flavor symmetric predictions are presented using the 
values for $F$\&$D$ which are mentioned in section III. Analytic
expressions which relate these parameters to the $g_A/g_V$ ratios
may {\it e.g.} be found in table I of \protect\cite{Pa90}.}
~\vskip0.01cm
\begin{tabular}{ c ||c | c | c | c | c | c}
&$n\to p$ & $\Lambda\to p$ & $\Sigma\to n$ & $\Xi\to\Lambda$ &
$\Xi\to\Sigma$ & $\Sigma\to\Lambda$\\
\hline
emp.&$1.258$ & $0.718\pm0.015$ & $0.340\pm0.017$ & $0.25\pm0.05$ &
$1.287\pm0.158$ & $0.61\pm 0.02$\\
$F$\&$D$&$1.258$ & $0.725\pm0.009$ & $0.339\pm0.026$ & $0.19\pm0.02$ &
$1.258=g_A$ & $0.65\pm0.01$
\end{tabular}
\end{table}

Our treatment of flavor symmetry breaking is based on the Skyrme model
approach to describe baryons as solitons in an effective meson theory.
In this type of models baryon states are obtained by quantizing 
the large amplitude fluctuations (zero modes) of the soliton. In the 
proceeding section we will briefly review the quantization procedure with
the inclusion of flavor symmetry breaking. This approach is of great 
fundamental interest not only for the large $N_C$ treatment of QCD but 
certainly also in its own right; especially because this approach has 
been (at least quantitatively) successful in understanding the proton 
spin problem \cite{Br88,Jo90,We96}. Furthermore, in the framework of 
quantizing the soliton the study of flavor symmetry breaking in these models 
is very interesting. This is even more the case as some of the difficulties 
encountered earlier (such as the overall scale in the predicted baryon 
mass differences~\cite{Pr83,Ch85} or the unexpectedly large strangeness 
contribution to nucleon matrix elements~\cite{Do86,Br88}) have been 
largely understood and solved~\cite{Ya88,We96,Pa89}. It is thus appealing 
to also study the $\Lambda$ axial charges in such a framework, especially 
because they might be accessible experimentally~\cite{Bo98,Lu95,El96}.
Here we will focus on a description with the symmetry breaking 
mainly residing in the baryon wave--functions, including important 
higher order contributions. The order parameter is the {\it strange} 
current quark mass, $m_s$. In the effective meson Lagrangian it
emerges via the meson properties, {\it e.g.} 
$m_K^2-m_\pi^2={\cal O}(m_s)$. Sizable 
deviations from flavor symmetric (octet) wave--functions are needed in 
the chiral soliton approach to account for the pattern of the baryon 
mass--splittings \cite{We96}. The proposed picture implies that the 
{\it strange} quark component in the sea is suppressed, a scenario which
has also been considered in ref \cite{Ka99}. On the other hand we will 
assume that the current operators, from which the charges are computed, 
are dominated by their flavor symmetric components. We will find that 
the proposed approach approximately reproduces the data with no (or minimal) 
explicit symmetry breaking in the axial charge operator. The present 
studies represent a refinement of some earlier calculations 
as we now include contributions to the axial charge operator which 
were omitted in ref \cite{Pa90} and are subleading in the $1/N_C$
counting. A systematic expansion in $1/N_C$ would
also require a careful treatment of the allowed representations in 
flavor space for the baryon wave--functions. We do not attempt such an
expansion but rather assume the physical value $N_C=3$. In addition we 
present the results obtained from a complete calculation in a realistic 
vector meson soliton model. That calculation supports the suggested picture.

\section{Symmetry Breaking in the Baryon Wave--Functions} 

Here we briefly review the energy eigenvalue problem for the 
low--lying $\frac{1}{2}^+$ and $\frac{3}{2}^+$ baryons as it arises 
in the collective coordinate treatment of chiral soliton models. 
This approach was initiated in ref~\cite{Ya88}. In the model
framework it leads to exact eigenstates for an arbitrary strength 
of the flavor symmetry breaking. The collective coordinates for flavor 
rotations are introduced via
\be
U(\vec{r},t)=A(t)U_0(\vec{r})A^\dagger(t)\, ,\qquad 
A(t)\in SU(3)\, .
\label{collcord}
\ee
$U_0(\vec{r})$ describes the soliton field configuration embedded
in the isospin subgroup of flavor $SU(3)$. A prototype model Largangian
for the chiral field $U(\vec{r},t)$ would consist of the Skyrme model 
supplemented by the Wess--Zumino--Witten terms as well as suitable 
symmetry breaking pieces. In the action notation it reads
\be
\Gamma=
\int d^4 x \Big\{ {f^2_\pi \over 4} 
\Tr\left[ \partial_\mu U (\partial^\mu U)^\dagger \right]
+
 {1\over{32 \skp^2}}
 \Tr\left[ [U^\dagger \partial_\mu U , 
U^\dagger \partial_\nu U]^2\right] \Big\} + \Gamma_{WZ}
+ \Gamma_{SB}\, .
\ee
Here $f_\pi$ is the pion decay constant and $\epsilon$ is 
the dimensionless Skyrme parameter. 
$\Gamma_{WZ}$ is the Wess-Zumino action \cite{Wit83}:
\be
\Gamma_{WZ} &=& - \frac{i N_C}{240 \pi^2}
\int_{M_5} d^5x \ \epsilon^{\mu\nu\rho\sigma\tau}\
\Tr[ L_\mu L_\nu L_\rho L_\sigma L_\tau] \quad {\rm with} 
\quad \partial M_5 = M_4\, .
\label{WZ}
\ee
Here we have used  $L_\mu = U^\dagger \partial_\mu U$. The flavor 
symmetry breaking terms are contained in 
\be
\Gamma_{SB} & = &\int d^4x \left\{
 { f_\pi^2 m_\pi^2 + 2 f_K^2 m_K^2 \over{12} }
 \Tr \left[ U + U^\dagger - 2 \right] 
+{f_\pi^2 m_\pi^2 - f_K^2 m_K^2 \over{2\sqrt{3}}}
\Tr \left[ \lambda_8 \left( U + U^\dagger \right) \right] \right.
\nonumber \\
& & \qquad
 \left.
+ { f_K^2 - f_\pi^2\over{4} }
\Tr \left[ {\hat S} \left(
U (\partial_\mu U)^\dagger \partial^\mu U + 
U^\dagger \partial_\mu U (\partial^\mu U)^\dagger \right)
\right] \right\} \, ,
\label{sb}
\ee
\noindent
where ${\hat S}={\rm diag}(0,0,1)$ is the strangeness projector.
It should be emphasized that many of the arguments presented below 
apply to more general chiral Lagrangians, though.

An appropriate parameterization
of the collective coordinates in terms of eight ``Euler--angles'' is 
given by
\be
A=D_2(\hat{I})\,{\rm e}^{-i\nu\lambda_4}D_2(\hat{R})\,
{\rm e}^{-i(\rho/\sqrt{3})\lambda_8}\ ,
\label{Apara}
\ee
where $D_2$ denote rotation matrices of three Euler--angles for 
each, rotations in isospace~($\hat{I}$) and 
coordinate--space~($\hat{R}$). Substituting
this configuration into the model Lagrangian yields 
upon canonical quantization the Hamiltonian for the
collective coordinates~$A$: 
\be 
H=H_{\rm s}+\fract{3}{4}\, \gamma\, {\rm sin}^2\nu\, .
\label{Hskyrme}
\ee
The symmetric piece of this collective Hamiltonian only contains 
Casimir operators and may be expressed in terms of the $SU(3)$--right 
generators $R_a\, ,$ with $[A,R_a]=(1/2)A\lambda_a\, ,$ 
where $a=1,\ldots,8\, :$
\be
H_{\rm s}=M_{\rm cl}+\frac{1}{2\alpha^2}\sum_{i=1}^3 R_i^2
+\frac{1}{2\beta^2}\sum_{\alpha=4}^7 R_\alpha^2\, .
\label{Hsym}
\ee
$M_{\rm cl},\alpha^2,\beta^2$ and $\gamma$ are functionals of the 
soliton, $U_0(\vec{r})$. The field theoretical problem has been 
transformed into a quantum mechanical problem for the collective 
coordinates which are parameterized by the `Euler--angles'~(\ref{Apara}).
The symmetry breaking term in the 
Hamiltonian~(\ref{Hskyrme}) depends on only one of the eight 
`Euler--angles'. This suggests the following parameterization 
of the baryon eigenfunctions \cite{Ya88},
\be
\Psi_{I,I_3,Y;J,J_3,Y_R}(A)=\frac{1}{\sqrt{N}}
\sum_{M_L,M_R}
D^{(I)*}_{I_3,M_L}(\hat{I})\,
f_{M_L,M_R}^{(I,Y;J,Y_R)}(\nu)\,{\rm e}^{iY_R\rho}\,
D^{(J)*}_{M_R,-J_3}(\hat{R})\, .
\label{Dpsi}
\ee
The unit baryon number sector constrains the right hypercharge to $Y_R=1$.
This constraint stems form $\Gamma_{WZ}$ and is valid for $N_C=3$.
The flavor hypercharge quantum number emerges via the constraint
$Y-Y_R=2(M_L-M_R)$ for the intrinsic (iso--)spin projections 
$M_L$ and $M_R$.

The generators $R_a$ can be expressed in terms of derivatives with respect
to the `Euler--angles'. Then the eigenvalue problem $H\Psi=\epsilon\Psi$
reduces to sets of ordinary second order differential equations for the
isoscalar functions $f_{M_L,M_R}^{(I,Y;J,Y_R)}(\nu)$. The product 
$\omega^2=\frac{3}{2}\gamma\beta^2$ appears as a continuous parameter in 
the eigenvalue equation. Hence the eigenfunctions~(\ref{Dpsi}) 
parametrically depend on $\omega^2$ which is thus interpreted as the 
effective strength of the flavor symmetry breaking. A value in the range
$5\lapeq\omega^2\lapeq8$ is required to obtain reasonable agreement with 
the empirical mass differences for the $\frac{1}{2}^+$ and $\frac{3}{2}^+$ 
baryons~\cite{We96}. In particular, reproducting the observed spacing
$(M_\Lambda-M_N):(M_\Sigma-M_\Lambda):(M_\Xi-M_\Sigma)=1:0.43:0.69$
demands a sizable $\omega^2$ since a leading order treatment of the 
eigenvalue equation~(\ref{Hskyrme}) incorrectly yields $1:1:\fract{1}{2}$.
In the exact treatment we get significantly closer to the empirical
values, {\it e.g.} for $\omega^2=6.0$ and $\omega^2=8.0$ we find the ratios 
$1:0.69:0.70$ and $1:0.61:0.77$, respectively\footnote{One might want
to add other symmetry breaking operators to (\ref{Hskyrme}) but it
should be reminded that they are of higher order in $1/N_C$.}.

The symmetry breaking piece in eq~(\ref{Hskyrme}) has non--zero
matrix elements when sandwiched between baryon states that differ
only by their respective $SU(3)$ representation. Hence the exact as
well as the perturbative treatments lead to baryon states which are
not pure octet (or decuplet) states. Rather they have admxitures of
states that are members of higher dimensional $SU(3)$ representations
but otherwise have identical quantum numbers. Baryons in 
these representations cannot be constructed as three quark states,
rather additional quark--antiquark pairs are required. Hence such 
admxitures to the octet (or decuplet) baryon wave--functions can be
interpreted as an effective parameterization of the meson cloud.
In general those admixtures reduce the baryon matrix elements associated 
with operators like $\bar{s}s$\cite{We96}\footnote{For example, the
normalized nucleon matrix element $\langle N|\bar{s}s|N\rangle/
\langle N|\bar{u}u+\bar{d}d+\bar{s}s|N\rangle$ reduces from 
23\% for the pure octet wave--function to 17\% at $\omega^2=5.0$.}.
Hence the meson cloud dominantly consists of pions. This merely
reflects the fact that due to the smaller masses the non--strange 
degrees of freedom are easier to excite than those related to
strange quarks. Here we are interested in the consequences that
arise from the exact treatment of the collective coordinates. We
therefore compare the admxitures of states from higher dimensional
representations as they result from the exact calculation outlined 
above with those obtained in the first order approximation.
As an example we list the admixture of the nucleon wave--functions with
states carrying nucleon quantum numbers dwelling in higher dimensional
$SU(3)$ representations in table \ref{amplitude}. 
\begin{table}[htb]
\caption{\label{amplitude}\sf The amplitude of various $SU(3)$ 
representations in the nucleon wave-functions. Presented are the 
exact and the first order results. In the exact treatment representations 
of higher dimensions than the ${\bf \bar{35}}$ also have non--vanishing 
amplitudes but they are not shown here.}
~\vskip0.01cm
\begin{tabular}{r| c c c c | c c c c}
& \multicolumn{4}{c|}{exact} &\multicolumn{4}{c}{first order}\\
\hline
$\omega^2$ & 
${\bf 8}$ & ${\bf \bar{10}}$ & ${\bf 27}$ & ${\bf \bar{35}}$ &
${\bf 8}$ & ${\bf \bar{10}}$ & ${\bf 27}$ & ${\bf \bar{35}}$ \\
\hline
4.0 & 0.977 & 0.170 & 0.128 & 0.018
    & 1.000 & 0.200 & 0.130 & 0.000 \\
6.0 & 0.955 & 0.231 & 0.184 & 0.036
    & 1.000 & 0.300 & 0.196 & 0.000 \\
8.0 & 0.927 & 0.278 & 0.233 & 0.056
    & 1.000 & 0.400 & 0.261 & 0.000 \\
10.0& 0.904 & 0.314 & 0.276 & 0.077
    & 1.000 & 0.500 & 0.326 & 0.000
\end{tabular}
\end{table}
We observe that in the relevant range for $\omega^2$ the first order 
approximation has only limited validity. In particular the 
$\bar{\bf 10}$--amplitude is overestimated by this approximation.

The feature that the effective symmetry breaking parameter also
contains the moment of inertia, $\beta^2$ for rotations into strangeness
direction allows the possibility that the symmetry breaking in the
wave--functions, which is measured by $\omega^2$, to be large albeit
the explicit symmetry breaking, measured by $\gamma$, is not (and 
{\it vice versa}). Furthermore this allows for the scenario of having 
large deviations from flavor symmetric wave--functions without even having 
symmetry breaking components in the current operators since almost all 
symmetry breaking can eventually be included in non--derivative terms 
of $\Gamma_{SB}$ which do not contribute to currents. In the next section 
we will study whether such a picture can be consistent with the 
observations on hyperon beta--decays. These decays are well parameterized 
by the Cabibbo scheme \cite{Ca63} which is obtained by applying the 
Wigner--Eckart theorem to the $SU(3)$ symmetric baryon octet wave--functions.

\section{Charge Operators} 

In this section we present an investigation based on the 
covariance in the collective coordinate approach. In this context
it is not necessary to detail the model Lagrangian. In the 
proceeding section we will nevertheless present an analysis 
which utilizes a specific Lagrangian as an example. It will be 
found that this example essentially verifies the results 
obtained from the covariant treatment.

Recently a chiral soliton model motivated analysis of
the axial charges of the hyperons has been performed \cite{Ki00}. Up
to linear order in the {\it strange} current quark mass, $m_s$
(next--to--leading order in flavor symmetry breaking) all operators
for the respective matrix elements were collected. Their coefficients were
determined from known data on hyperon beta--decay\footnote{Note that the 
standard definition for this decay parameter differs from that in 
ref~\protect\cite{Ki00} by a factor~$\sqrt{6}/2$.}. A model result was used
to relate octet and singlet currents because they are not related by group
theory. Then the polarization for the {\it non--strange} quarks in the
$\Lambda$ was predicted to be small, $\Delta U_\Lambda=-0.02\pm0.17$ in
contrast to $\Delta S_\Lambda=1.21\pm0.54$; with errors of the $\Xi$ decay
data penetrating through this analysis. Some of the results (for the central
values) raise questions in view of the study representing a perturbation
expansion in flavor symmetry breaking: The axial {\it singlet} matrix
element of the $\Lambda$, $\Delta\Sigma_\Lambda$, turned out to be about
twice as large as that of the nucleon, $\Delta\Sigma_N$. Also, the
${\cal O}(m_s)$ terms contributed almost 50\% to $\Delta S_\Lambda$. This
indicates that at this order the expansion has not converged (if it does
at all) or that in chiral soliton models the flavor symmetric point may
not be the most suitable one to expand about. This may be perceived
from the observation that in chiral soliton models the effect of the
derivative type symmetry breaking terms is mainly indirect. They provide 
the splitting between the various decay constants and thus significantly 
increase $\gamma$ because it is proportional 
to $f_K^2m_K^2-f_\pi^2m_\pi^2\approx 1.5f_\pi^2(m_K^2-m_\pi^2)$. 
Besides this indirect effect the derivative type symmetry breaking 
terms in (\ref{sb}) may be omitted. Whence there are no symmetry
breaking terms in current operators and the octet axial charge 
operator may be parameterized as
\be
\int d^3r A_i^{(a)} = c_1 D_{ai} - c_2 D_{a8}R_i
+c_3\sum_{\alpha,\beta=4}^7d_{i\alpha\beta}D_{a\alpha}R_\beta
\, ,\qquad a=1,\ldots,8\,,\quad i=1,2,3\, .
\label{axsym}
\ee
Under flavor transformations (parameterized by changes of the collective
coordinates) these charge operators behave like members of an octet.
The expression (\ref{axsym})
stems from subsituting the parameterization (\ref{collcord}) into the 
covariant expression of the current operator obtained from a model
Lagrangian using Noether's theorem and subsequently applying the 
quantization rules for the collective coordinates. Here we have 
furthermore introduced the adjoint representation of the collective rotations,
$D_{ab}=\frac{1}{2}{\rm tr}\left(\lambda_a A\lambda_b A^\dagger\right)$.
In principle, the constants $c_n, n=1,2,3$, are functionals of the
soliton which can be computed within the adopted model.
The $c_2$--term originates solely from the abnormal parity
terms in the action, {\it e.g.} $\Gamma_{WZ}$, while 
the $c_3$--term additionally acquires contributions from field components 
which are induced by the collective rotations. Both, $c_2$ and $c_3$
are subleading in $1/N_C$ as the appearance of the generators,
$R_a$ suggests. A well--known problem of many chiral soliton models is 
the too small prediction for the axial charge of the nucleon, $g_A$ when 
the constants $c_n$ are computed using the soliton solution. In this
section we will not address that problem but rather use the 
empirical value $g_A=1.258$ as an input to determine the $c_n$.
That is, we consider the constants $c_n$ as free parameters, alike
$F$\&$D$ in the Cabibbo scheme.

\begin{figure}[htb]
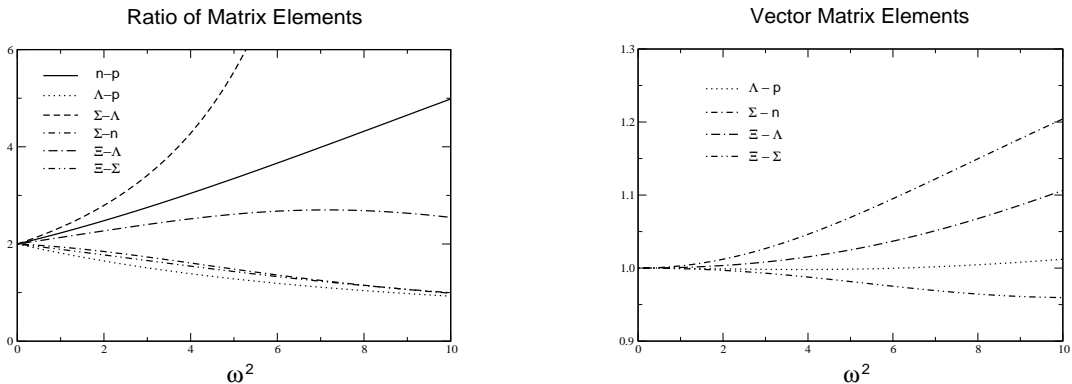

\centerline{
\epsfig{figure=ratio.eps,height=6.0cm,width=5cm,angle=270}
\hskip2cm
\epsfig{figure=vector.eps,height=6.0cm,width=5cm,angle=270}}
~\vskip0.2cm
\caption{\label{fig_ratio}\sf The ratio of the matrix elements
$\langle B |D_{a3}| B^\prime\rangle/
\langle B |\sum_{\alpha,\beta=4}^7d_{3\alpha\beta}D_{a\alpha}R_\beta
| B^\prime\rangle$ for the relevant baryon states $B$ and $B^\prime$
as a function of the effective symmetry breaking parameter $\omega^2$.
The right panel shows the dependence of the vector matrix elements
on symmetry breaking (\protect\ref{vector}). They are normalized to 
the symmetric case.}
\end{figure}

It turns out that for pure octet wave--functions the matrix elements of
the operators multiplying the constants $c_1$ and $c_3$ have the same 
ratio $F/D=5/9$ while the operator associated with $c_2$ has $F/D=-5/3$. 
This suggests to put $c_1+c_3/2=-(3F+5D)/2$ and $c_2=(9F-5D)/\sqrt{3}$ with 
the empirical values $g_A=F+D=1.258$ and $F/D=0.575\pm0.016$, {\it i.~e.}
$c_1+c_3/2\approx-2.69$ and $c_2\approx0.09$. Of course, 
these relations are correct only for $\omega^2=0$. 
To see that the parameterization of the axial current matrix elements
in terms of $F$\&$D$ Clebsch--Gordan coefficients
becomes invalid already at moderate $\omega^2$ we consider the 
ratios $\langle B |D_{a3}| B^\prime\rangle/
\langle B |\sum_{\alpha,\beta=4}^7d_{3\alpha\beta}D_{a\alpha}R_\beta
| B^\prime\rangle$ in figure \ref{fig_ratio}.
The fact that the operators $D_{a3}$ and
$\sum_{\alpha,\beta=4}^7d_{3\alpha\beta}D_{a\alpha}R_\beta$
have the same $F/D$ ratio is
reflected by all ratios assuming the same value when flavor
covariant wave--functions are used ($\omega^2=0$). However, we see
that already at moderate symmetry breaking the description of the
axial current matrix elements in terms of $F/D$ ratios
becomes inadequate as these operators evolve quite differently.
With these significant dependencies on the effective symmetry breaking
of matrix elements of the various operators contributing to the axial 
charges on $\omega^2$ it seems difficult to 
imagine that the empirical results for the hyperon decays, which are 
well described by the symmetric formulation, can be reasonably 
reproduced at realistic $\omega^2\gapeq5$. 

Before attempting such a fit we can get more 
insight into the relevance of the constants $c_n$ from the axial
singlet current. Although it is not related to the octet current
(\ref{axsym}) by group theoretical means, the fact that within the
collective coordinate approach we can
consider flavor symmetry breaking as a continuous parameter
provides further information. In the limit $\omega^2\to\infty$
(integrating out {\it strange} degrees of freedom)
the model should reduce to the two flavor formulation. In 
particular the strangeness contribution to the axial charge
of the nucleon should vanish in that limit. Noting that
$\langle N| D_{83}| N\rangle\to0$ and 
$\langle N| \sum_{\alpha,\beta=4}^7d_{3\alpha\beta}D_{8\alpha}R_\beta
| N\rangle\to0$ while $\langle N| D_{88}| N\rangle\to1$ for 
$\omega^2\to\infty$, we demand 
\be
\int d^3r A_i^{(0)}= -2\sqrt{3} c_2 R_i\quad i=1,2,3\,.
\label{singsym}
\ee
for the axial singlet current because it leads to the strangeness
projection
\be
\int d^3r A_i^{(s)}&=&\frac{1}{3}
\int d^3r\left( A_i^{(0)}-2\sqrt{3}A_i^{(8)}\right)
\nonumber \\
&=&\frac{-2}{\sqrt{3}}\left\{c_1D_{8i}+c_2(1-D_{88})R_i
+c_3\sum_{\alpha,\beta=4}^7d_{i\alpha\beta}D_{8\alpha}R_\beta\right\}\, .
\label{axstrange}
\ee
Actually all model calculations in the literature \cite{Pa92,Bl93} 
are consistent with this requirement. It is simply a consequence of 
embedding the soliton in the isospin subgroup of flavor $SU(3)$. The 
analysis of the famous proton spin puzzle yielding 
$\Delta\Sigma_N=\langle N|\int d^3r A_i^{(0)}|N\rangle=
0.20\pm0.10$ then suggests 
$c_2=0.12\pm0.06$ in agreement with the above estimate from the
flavor symmetric description of hyperon decays.

In order to completely describe the hyperon beta--decays we
also demand matrix elements of the vector charges. These are obtained 
from the operator
\be
\int d^3r V_0^{(a)} = \sum_{b=1}^8D_{ab}R_b=L_a,
\label{vector}
\ee
which introduces the left $SU(3)$ generators $L_a$. Again, this relation
is obtained by substituting (\ref{collcord}) into the covariant expression
for the vector current operator extracted from the model Lagrangian.
The relevant matrix elements are protected by the Ademollo--Gatto 
theorem~\cite{Ad64} stating that deviations from the $SU(3)$ relations 
start at order $(\omega^2)^2$. In the collective coordinate appraoch
this theorem is reproduced as the slope of these curves vanishes at 
$\omega^2=0$, {\it cf.} figure~\ref{fig_ratio}. Consequently, symmetry 
breaking in the vector currents is not only ignored in the Cabibbo scheme 
but also in the linear treatment of ref~\cite{Ki00}. However, for the 
strongly distorted wave--functions, which we are utilizing,
the deviations from the $SU(3)$ relations is sizable as is 
clearly shown in figure~\ref{fig_ratio}. Of course, we will take into
account these deviations when computing the vector charges. 

\begin{figure}[t]
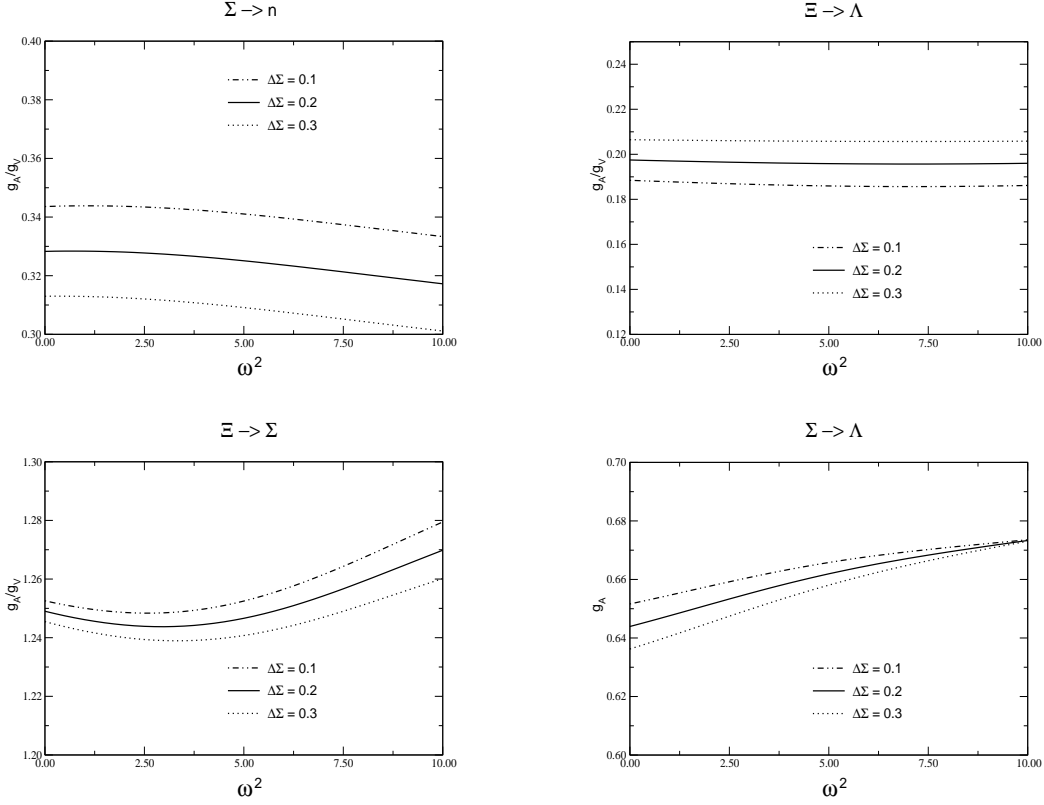

\centerline{
\epsfig{figure=nusi.eps,height=6.0cm,width=5cm,angle=270}
\hspace{1.5cm}
\epsfig{figure=chla.eps,height=6.0cm,width=5cm,angle=270}}
~\vskip0.05cm
\centerline{
\epsfig{figure=sich.eps,height=6.0cm,width=5cm,angle=270}
\hspace{1.5cm}
\epsfig{figure=sila.eps,height=6.0cm,width=5cm,angle=270}}
~\vskip0.3cm
\caption{\label{decay}\sf The predicted decay parameters for the 
hyperon beta--decays using $\omega^2_{\rm fix}=6.0$. 
The errors originating from those in $\Delta\Sigma_N$ are indicated.}
\end{figure}

We now attempt to determine the constants $c_n$ to reasonably 
fit the ratios $g_A/g_V$ for the hyperon beta--decays (only $g_A$ 
for $\Sigma^+\to\Lambda e^+\nu_e$). The values for $g_A$ and $g_V$ 
are obtained from the appropriate matrix elements of respectively 
the operators in eqs~(\ref{axsym}) and~(\ref{vector}), sandwiched 
between the eigenstates of the full Hamiltonian~(\ref{Hskyrme}).
We first have to fix a value, $\omega^2_{\rm fix}$ for which we want 
to obtain the best fit. We adopt the following strategy: we choose $c_2$ 
according the proton spin puzzle and subsequently determine $c_1$ and 
$c_3$ at $\omega^2_{\rm fix}=6.0$ such that the nucleon axial charge, 
$g_A$ and the $g_A/g_V$ ratio
for $\Lambda\to p e^-\bar{\nu}_e$ are reproduced. For example, setting
$\Delta\Sigma=0.2$ yields $c_1=-1.97$, $c_2=0.12$, and $c_3=-1.38$.
This is not too different from the above consideration in the 
symmetric case as $c_1+c_3/2=-2.66$. The matrix elements for
the $n\to p$ and  $\Lambda\to p$ transitions enter this determination
of the~$c_n$. The comparision with figure \ref{fig_ratio} tells us that 
the deviations from the symmetric limit have turned out unexpectedly small. 
We are now left with predictions not only for the decay parameters of 
the other decay processes but we can also study the variation with symmetry 
breaking of all relevant decays. This is shown in figure~\ref{decay}.
Obviously the dependence on flavor symmetry breaking is very moderate, 
on the order of only a few percent. In view of the model being 
an approximation this dependence may be considered irrelevant
and the results can be viewed as being in reasonable agreement 
with the empirical data, {\it cf.} table \ref{empirical}. 
The observed independence of $\omega^2$ shows that these
predictions are not sensitive to the choice of $\omega^2_{\rm fix}$.
In addition, since we observe this approximate independence 
of $\omega^2$, we essentially have a two parameter ($c_1$ and $c_3$,
$c_2$ is fixed from $\Delta\Sigma_N$) fit of the hyperon beta--decays. 
This is alike the $F$\&$D$ parameterization in the Cabibbo scheme.
We remark that the two transitions, $n\to p$
and $\Lambda\to p$, which are not shown in figure~\ref{decay},
exhibit a similar neglegible dependence on $\omega^2$ and, by 
construction, they match the empirical data at $\omega^2=6.0$.
It should be noted that the use of the exact eigenfunctions 
of (\ref{Hskyrme}), which leads to the non--linear behavior is important 
in this regard. A linearized version (in $\omega^2$) would not have 
necessarily yielded this result. In particular a first
order description would fail for the process $\Xi\to\Sigma$, 
for which $g_A/g_V$ is a non--monotonous function of $\omega^2$.
Comparing the results shown in figure \ref{decay} with the data in 
table \ref{empirical} we see that the calculation using the strongly 
distorted wave--functions agrees approximately as well with the empirical 
data as the flavor symmetric $F$\&$D$ fit. 

We also observe that the singlet current does not get modified. Hence we 
have the simple relation
\be
\Delta\Sigma_N=\Delta\Sigma_\Lambda
\label{lasing}
\ee
for all values of $\omega^2$.

\begin{figure}[t]
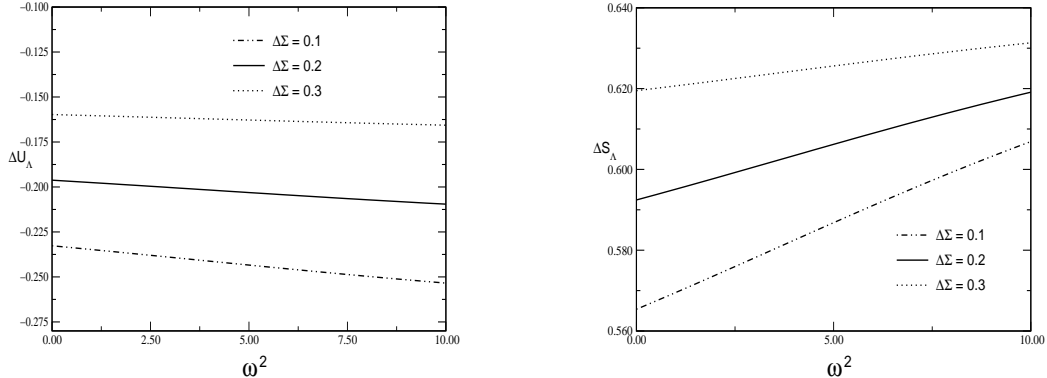

\centerline{
\epsfig{figure=h1la.eps,height=6.0cm,width=5cm,angle=270}
\hspace{1.5cm}
\epsfig{figure=h3la.eps,height=6.0cm,width=5cm,angle=270}}
~\vskip0.3cm
\caption{\label{laxial}\sf The contributions of the {\it non--strange}
(left panel) and {\it strange} (right panel) degrees of freedom
to the axial charge of the $\Lambda$. Again we used
$\omega^2_{\rm fix}=6.0$.}
\end{figure}

In figure \ref{laxial} we display the flavor components of the axial 
charge of the $\Lambda$ hyperon. We see that also the various contributions 
to the axial charge of the $\Lambda$ only exhibit a moderate dependence 
on $\omega^2$. The {\it non--strange} component, 
$\Delta U_\Lambda=\Delta D_\Lambda$ slightly increases in magnitude. 
The {\it strange} quark piece, $\Delta S_\Lambda$ then grows with 
symmetry breaking since we keep $\Delta\Sigma_\Lambda$ fixed. It should 
be remarked that the results shown in figure~\ref{laxial} agree nicely 
with an $SU(3)$ analysis applied to the data \cite{Ja96,Bu93,Bo98}: 
$\Delta U_\Lambda=\Delta D_\Lambda\approx-0.20$ and 
$\Delta S_\Lambda\approx0.60$. Finally we remark that the observed
independence on the symmetry breaking does not occur for all
matrix elements of the axial current. An interesting counter--example
is the {\it strange} quark component in the nucleon, $\Delta S_N$. 
For $\Delta\Sigma=0.2$, say, it is significant at zero symmetry
breaking, $\Delta S_N=-0.131$ while it decreases (in magnitude) to 
$\Delta S_N=-0.085$ at $\omega^2=6.0$.

Within this class of models the order of the expansion in symmetry 
breaking is measured by differences like $\omega^2\sim m_K^2-m_\pi^2$, 
$f_K^2-f_\pi^2$ or $m_{K^*}^2-m_\rho^2$ which are linear in the 
{\it strange} current quark mass $m_s$. In an systematic expansion one 
could add symmetry breaking components to the currents to eliminate the 
small deviations from the empirical data. As these corrections are 
potentially small it might well be that this could be accomplished by a 
single operator of ${\cal O}(m_s^2)$ or even higher. In turn this would 
make the approach quite unpredictable. In addition the errors in the 
empirical data ({\it cf.} table~\ref{empirical}) may penetrate to the 
fitted coefficients~$c_n$. It seems thus more appropriate 
to revert to realistic models in which we can calculate the coefficients 
of the next--to--leading order terms and which have been tested at 
other instances.

\section{Spin Content of the $\Lambda$ in a Realistic Model}

We consider a realistic soliton model which contains pseudoscalar 
and vector meson fields. It has been established for two flavors in 
ref \cite{Ja88} and been extended to three flavors in ref~\cite{Pa92} 
where it has been shown to fairly describe the parameters of hyperon 
beta--decay ({\it cf.} table~4 in ref~\cite{Pa92}). 

Starting point is a three--flavor chirally invariant theory for 
pseudoscalar and vector mesons. The model Lagrangian contains terms which 
involve the Levi--Cevita tensor $\epsilon_{\mu\nu\rho\sigma}$, to 
accommodate processes like $\omega\rightarrow3\pi$~\cite{Ka84}. 
These terms contribute to $c_2$ and $c_3$.  A minimal set of 
symmetry breaking terms is included \cite{Ja89} to 
account for different masses and decay constants. This effective 
theory contains topologically non--trivial static solutions, which 
are constructed by imposing {\it ans\"atze} in the isospin subgroup
\begin{eqnarray}
\xi(\vec{r})={\rm exp}\left(\fract{i}{2}\,
\hat{\vec{r}}\cdot\vec{\tau}F(r)\right),
\quad
\omega_0(\vec{r})=\omega(r)
\quad {\rm and} \quad
\rho_{i,a}(\vec{r})=\frac{G(r)}{r}\,\epsilon_{ija}\hat r_j\ ,
\label{solan}
\end{eqnarray}
while all other field components vanish classically. Here 
$\xi={\rm exp}\left(i\vec{\pi}\cdot
\vec{\tau}/2f_\pi\right)$ refers to the non--linear 
realization of the pion fields. The radial functions are determined 
by extremizing the static energy functional subject to boundary 
conditions appropriate to unit baryon number. Introducing collective
coordinates for this configuration induces field components which are 
classically absent. From this, eight real radial functions emerge. They 
solve inhomogeneous linear differential equations with the soliton 
profiles (\ref{solan}) acting as sources. In regard of the discussion in 
the preceding section it is interesting to note that despite of strong 
symmetry breaking in the baryon wave--functions the model predictions
for the magnetic moments approximately obey the respective $SU(3)$
relations~\cite{Pa92}.

Covariant expressions for the (axial--)vector currents are obtained
by introducing appropriate sources. Substituting the above described 
{\it ans\"atze} and applying the quantization rules for the collective 
coordinates yields the charges as linear combinations of functionals, 
$c_n[F,\omega,G,...]$ of the meson profile functions and operators in 
the space of the collective coordinates $A$. In this model the derivative 
type symmetry breaking terms add symmetry breaking pieces to the axial 
charge operator,
\be
\delta A_i^{(a)}=c_4 D_{a8}D_{8i}+
c_5 \sum_{\alpha,\beta=4}^7d_{i\alpha\beta}D_{a\alpha}D_{8\beta}+
c_6 D_{ai}(D_{88}-1)\quad {\rm and}\quad
\delta A_i^{(0)}= 2\sqrt{3}\,c_4D_{8i}\, .
\label{axnonsym}
\ee
The identical coefficient $c_4$ in the octet and singlet currents 
arises from the model calculations, it is not demanded
by the above mentioned consistency condition of having vanishing 
strangeness contribution in the nucleon for $\omega^2\to\infty$ since 
we find $\langle N| D_{88}D_{83} | N\rangle\to 0$ as well as 
$\langle N| D_{83} | N\rangle\to 0$.

Once the model parameters are agreed on, the coefficients 
$c_1,\ldots,c_6$ are uniquely determined as are the parameters in the 
collective Hamiltonian, which in this model is more involved than
eq~(\ref{Hskyrme}). Thus the baryon wave--functions as well as the 
current operators are fixed and all relevant decay parameters can be 
computed. Unfortunately the model parameters cannot be completely 
determined in the meson sector~\cite{Ja88}. We use the remaining 
freedom to accommodate baryon properties in three different ways as
shown in table \ref{realistic}. The set denoted by `b.f.' refers 
to an overall best fit to the spectrum of the low--lying spin 
$\frac{1}{2}$ and spin $\frac{3}{2}$ baryons.
It predicts the axial charge somewhat on the low side, $g_A=0.88$.
The set named `mag.mom.' labels a set of parameters which yields 
magnetic moments which are close to the respective empirical data
(with $g_A=0.98$) and finally the set labeled `$g_A$' reproduces the 
axial charge of the nucleon and also reasonably accounts for hyperon
beta--decay \cite{Pa92}.
\begin{table}[htb]
\caption{\label{realistic}\sf Spin content of the $\Lambda$ in the
realistic vector meson model. For comparison the nucleon 
results are also given. Three sets of model parameters
are considered, see text.}
~\vskip0.01cm
\begin{tabular}{ c || c | c |c || c | c | c | c}
& \multicolumn{3}{c||}{$\Lambda$} &
\multicolumn{4}{c}{$N$}\\
\hline
& $\Delta U = \Delta D$ & $\Delta S$ & $\Delta\Sigma$ &
 $\Delta U$ & $\Delta D$ & $\Delta S$ & $\Delta\Sigma$\\
\hline
b.f.
&$-0.155$&$0.567$&$0.256$&$0.603$&$-0.279$&$-0.034$&$0.291$\\
mag. mom. 
&$-0.166$&$0.570$&$0.238$&$0.636$&$-0.341$&$-0.030$&$0.265$\\
$g_A$ 
&$-0.164$&$0.562$&$0.233$&$0.748$&$-0.476$&$-0.016$&$0.256$\\
\end{tabular}
\end{table}
For all three sets the effective symmetry breaking is sizable,
$\omega^2\approx10$. However, its effect is somewhat mitigated
by additional symmetry breaking terms 
($\sim \sum_{i=1}^3D_{8i}R_i$, 
$\sum_{\alpha=4}^7D_{8\alpha}R_\alpha$) in the collective
Hamiltonian~(\ref{Hskyrme}).
We observe that in particular the predictions for the 
axial properties of the $\Lambda$ are quite insensitive to 
the model parameters. The variation of the model parameters
only seems to influence the isovector part of the axial charge
operator. Surprisingly the singlet matrix element of the 
$\Lambda$ hyperon is smaller than that of the nucleon, although 
this effect is tiny. As this difference emerges solely from the
$c_4$ term this ordering is a reflection of $c_4$ being positive
in this model. It should be noted that in other models $c_4$ is
predicted to be negative~\cite{We92}, although small in magnitude 
as well; suggesting that $\Delta\Sigma_\Lambda\approx\Delta\Sigma_N$ 
in general.

Similar to the fit of the previous section the full model calculation
predicts sizable polarizations of the {\it up} and {\it down} quarks
in the $\Lambda$ which are slightly smaller in magnitude but nevertheless
comparable to those obtained from the $SU(3)$ symmetric analysis. One 
wonders whether the significant {\it up}--quark polarization 
$\Delta U_\Lambda=\Delta D_\Lambda\approx -0.16$ has an experimental
signature. The Gribov--Lipatov reciprocity relation~\cite{Gr71}
suggests that the quark fragmentation functions follow the quark 
distribution functions. Although there is no direct connection this 
hypothesis may nevertheless serve as an estimate~\cite{Br97,Bo99}. As the 
integrated polarized distribution functions are just $\Delta Q_B$ (for 
quark $Q$ inside baryon $B$) it is suggestive that the predicted value 
for $\Delta U_\Lambda$ goes along with a significant {\it up}--quark
fragmentation function for the $\Lambda$. In electroproduction the 
individual quark contributions to oberservables are weighted by the 
square of the respective charges. For the $\Lambda$ this elevates
the {\it non--strange} contribution by a factor five. Equations~(2)--(4) 
in ref~\cite{Ja96} give the polarization of a $\Lambda$ that is produced 
in the current fragmentation region of deep inelastic electron--proton 
scattering. In addition to the ${\rm (charge)}^2$ factor the 
{\it strange}--quark contribution is suppressed in such processes as 
its distribution in the nucleon is presumably small. In essence the 
significant negative prediction for $\Delta U_\Lambda$ should result in 
a negative polarization of $\Lambda$'s produced in that a reaction. In 
view of the self--analyzing decay $\Lambda\to p \pi^-$ this should be 
detectable. However, it is subject to the (reasonable) assumption that 
fragmentation and distribution functions are closely related.

\section{Conclusions}

In the collective coordinate approach to chiral solitons large deviations 
from flavor symmetric (octet) wave--functions are required to accommodate 
the observed pattern of the baryon mass--splitting. Especially, 
contributions which arise beyond next--to--leading order in the effective
symmetry breaking are needed for this purpose. In the QCD language these
are of the order $m_s^2$ or higher. In this report we have suggested a 
picture for the axial charges of the low--lying~$\frac{1}{2}^+$~baryons 
which manages to reasonably reproduce the empirical data without 
introducing (significant) flavor symmetry breaking components in the 
corresponding operators. Rather, the sizable symmetry breaking resides 
almost completely in the baryon wave--functions. This scenario is 
especially motivated by the Yabu--Ando treatment of the Skyrme model which 
has the major symmetry breaking components in the potential part of the 
action and thus no (or only minor) symmetry breaking pieces in the 
current operators. The empirical data for these decay parameters are as 
reasonably reproduced as in the Cabibbo scheme of hyperon beta--decay. 
Repeatedly we emphasize that the present picture is not a re--application 
of the Cabibbo scheme since in the present calculation the `octet' baryon 
wave--functions have significant admixture of higher dimensional 
representations ({\it cf.} table~\ref{amplitude}). Furthermore the 
individual matrix elements which enter this calculation may strongly 
vary with the effective symmetry breaking (or {\it strange} current 
quark mass), {\it cf.} figure~\ref{fig_ratio}; only
when combining them to the full $g_A/g_V$ ratios the strong dependence 
on the strength of symmetry breaking cancels.

In the present treatment we may consider symmetry breaking as
a continuous parameter. Taking this parameter to be infinitely large
the two flavor model must be retrieved for the nucleon. This consistency 
condition relates coefficients in the axial singlet current operator to 
the respective octet components, which are not otherwise related to each 
other by group theory. In turn we are enabled to completely
disentangle the quark flavor components of the axial charge. 
It results in sizable {\it up} and {\it down}
quark polarizations in the $\Lambda$. Again, a picture emerged which,
after some cancellations, agrees with that of the flavor symmetric
treatment for known data. These results were obtained utilizing a 
parameterization of a charge operator which did not contain
any symmetry breaking component. 

We have also considered a realistic model, wherein the parameters 
entering the charge operators are actually predicted. These operators 
contain non--vanishing symmetry breaking pieces, whose matrix elements 
are, however, small. Essentially this model calculation confirmed the 
results obtained in the parametrically treatment.

\subsection*{Acknowledgments}

The author would like to thank G.~R.~Goldstein, R.~L.~Jaffe and 
J.~Schechter for helpful conversations and useful references.

This work is supported in part by funds provided by
the U.S. Department of Energy (D.O.E.) under cooperative research agreement
\#DF-FC02-94ER40818 and the Deutsche Forschungsgemeinschaft (DFG) under
contract We 1254/3-1.


\begin{thebibliography}{99}
\bibitem{Ja96}
R. L. Jaffe, Phys. Rev. {\bf D54} (1996) 6581.
\bibitem{Bu93}
M. Burkardt and R. L. Jaffe, Phys. Rev. Lett. {\bf 70} (1993) 2537.
\bibitem{Fl97}
D. de Florian, M. Stratmann, and W. Vogelsang, hep--ph/9710410:
{\it Polarized Lambda Production at HERA}.
\bibitem{Bo98}
C. Boros and L. Zuo--tang, Phys. Rev. {\bf D57} (1998) 4491;
and references therein.
\bibitem{Ca63}
N. Cabibbo, Phys. Rev. Lett. {\bf 10} (1963) 531.
\bibitem{Fl98} R. Flores--Mendieta, E. Jenkins, and A. V. Manohar,
Phys. Rev. {\bf D58} (1998) 094028.
\bibitem{DATA}
C. Caso {\it et al.}, (Particle Data Group), Eur. Phys. J. {\bf C3}
(1998) 1,
M. Bourquin {\it et al.}, Z. Phys. {\bf C12} (1982) 307;
{\bf C21} (1983) 1.
\bibitem{Pa90}
N. W. Park, J. Schechter, and H. Weigel,
Phys. Rev. {\bf D41} (1990) 2836.
\bibitem{Br88}
S.~Brodsky, J.~Ellis, and M.~Karliner,
Phys. Lett. {\bf B206} (1988) 309.
\bibitem{Jo90}
R.~Johnson, N.~W. Park, J.~Schechter, V.~Soni, and H.~Weigel,
Phys. Rev. {\bf D42} (1990) 2998.
\bibitem{We96}
H. Weigel, Int. J. Mod. Phys. {\bf A11} (1996) 2419;
J. Schechter and H. Weigel, hep--ph/9907554:
{\it The Skyrme Model for Baryons}.
\bibitem{Pr83}
M.~Prasza{\l}owicz,
Phys. Lett. {\bf 158B} (1983) 264.
\bibitem{Ch85}
M.~Chemtob,
Nucl. Phys. {\bf B256} (1985) 600.
\bibitem{Do86}
J.~Donoghue and C.~R. Nappi,
Phys. Lett. {\bf 168B} (1986) 105.
\bibitem{Ya88}
H. Yabu and K. Ando,
Nucl. Phys. {\bf B301}, 601 (1988).
\bibitem{Pa89}
N.~W. Park, J.~Schechter, and H.~Weigel,
Phys. Lett. {\bf B228} (1989) 420.
\bibitem{Lu95}
W. Lu and B. Q. Ma, Phys. Lett. {\bf B357} (1995) 419.
\bibitem{El96}
J. Ellis, D. Kharzeev, and A. Kotzinan, Z. Phys. {\bf C69} (1996) 467.
\bibitem{Ka99}
M. Karliner and H. J. Lipkin,
Phys. Lett. {\bf B461} (1999) 280.
\bibitem{Wit83}
E. Witten,
Nucl. Phys. {\bf B223} (1983) 422;\
\"O.~Kaymakcalan, S.~Rajeev and J.~Schechter,
Phys. Rev. {\bf D30} (1984) 594.
\bibitem{Ki00}
H.--C. Kim, M. Prasza{\l}owicz, and K. Goeke,
Acta Phys. Polon. {\bf B31} (2000) 1767;
see also H.--C. Kim, M. Prasza{\l}owicz, and K. Goeke,
Phys. Rev. {\bf D61} (2000) 114006.
\bibitem{Pa92}
N. W. Park and H. Weigel,
Nucl. Phys. {\bf A541} (1992) 453.
\bibitem{Bl93}
A. Blotz, M. Prasza{\l}owicz, and K. Goeke,
Phys. Lett. {\bf B317} (1993) 195.
\bibitem{Ad64}
M. Ademollo and R. Gatto,
Phys. Rev. Lett. {\bf 13} (1964) 264.
\bibitem{Ja88}
P. Jain, R. Johnson, Ulf--G. Mei{\ss}ner, N. W. Park, and J. Schechter,
\newblock Phys. Rev. {\bf D37} (1988) 3252;\,
Ulf--G. Mei{\ss}ner, N. Kaiser, H. Weigel, and J. Schechter,
\newblock Phys. Rev. {\bf D39} (1989) 1956.
\bibitem{Ka84}
{\"O}. Kaymakcalan, S. Rajeev, and J. Schechter,
\newblock Phys. Rev. {\bf D30} (1984) 594.
\bibitem{Ja89}
P.~Jain, R.~Johnson, N.~W. Park, J.~Schechter, and H.~Weigel,
\newblock Phys. Rev. {\bf D 40} (1989) 855.
\bibitem{We92}
H. Weigel, R. Alkofer, and H. Reinhardt,
Nucl. Phys. {\bf B387} (1992) 638;\,
A. Blotz, D.~Diakonov, K. Goeke, N. W. Park, V. Petrov,
and P. V. Pobylitsa, Nucl. Phys. {\bf A555} (1993) 765.
\bibitem{Gr71}
V. N. Gribov and L. N. Lipatov, Phys. Lett. {\bf B37} (1971) 78.
\bibitem{Br97}
S. J. Brodsky and B.--Q. Ma, Phys. Lett. {\bf 392} (1997) 542.
\bibitem{Bo99}
C. Boros and A. W. Thomas, Phys. Rev. {\bf D60} (1999) 074071.
\end{thebibliography}
\end{document}